\documentclass[twocolumn,twoside,slac_two]{revtex4-1}
\usepackage{graphicx}
\usepackage{fancyhdr}
\usepackage{color}
\definecolor{mylinkcol}{rgb}{0.1, 0.1, 0.5}
\definecolor{mycitecol}{rgb}{0.1, 0.45, 0.05}
\definecolor{myurlcol}{rgb}{0.5, 0.1, 0.2}
\usepackage[breaklinks=true,colorlinks=true,linkcolor=mylinkcol,citecolor=mycitecol,urlcolor=myurlcol]{hyperref}
\usepackage{breakurl}
\begin{document}

\title{Extragalactic Jets from the TANAMI Sample as Seen by \textsl{Fermi}/LAT}

\author{Moritz B\"ock$^1$, Matthias Kadler$^{1,2,3}$, Gino Tosti$^{4}$,
Toby Burnett$^5$\\on behalf of the \textsl{Fermi}/LAT collaboration\\[10pt]
Roopesh Ojha$^6$, Cornelia M\"uller$^1$, J\"orn Wilms$^1$\\on
  behalf of the TANAMI collaboration\\~}

\affiliation{
  $^1$ Remeis-Observatory \& ECAP, FAU, 96049 Bamberg, Germany\\
  $^2$ CRESST/NASA GSFC, Greenbelt MD 20771, USA\\
  $^3$ USRA, Columbia MD 21044, USA\\
  $^4$ INFN/University of Perugia, 06123 Perugia, Italy\\
  $^5$ University of Washington, Seattle WA 98195, USA\\
  $^6$ NVI/United States Naval Observatory, Washington DC 20392, USA}

\begin{abstract}
The TANAMI program has been monitoring the parsec-scale radio jets of
southern $\gamma$-ray bright AGN with VLBI techniques simultaneously with
\textsl{Fermi}/LAT monitoring of their $\gamma$-ray emission. Here we present the
$\gamma$-ray properties of the TANAMI sources based on an analysis of the
preliminary 1-year LAT source list. We present upper limits on the
$\gamma$-ray flux for TANAMI sources not detected by LAT.
\end{abstract}

\maketitle

\thispagestyle{fancy}

\section{Introduction}

Blazars are a subset of active galactic nuclei (AGN). They are variable,
radio-loud, and very luminous. Their emission often has a high degree of
polarization. The strong $\gamma$-ray variability and luminosity can be
explained by collimated jets consisting of charged particles moving at
relativistic velocities, which are orientated with a small angle to the
line of sight and thus Doppler boosted
\citep{Blandford1978,Maraschi1992}. This explanation is confirmed by the
fact that blazars typically exhibit apparent superluminal motion in the
inner radio-jet \citep{Lister2009_kinematics}.

The EGRET detector aboard the \textsl{Compton} Gamma Ray Observatory
\citep[\textsl{CGRO};][]{Thompson1993_egret} found that many blazars have
strong $\gamma$-ray emission. A successor of this mission was launched in
2008 June 11: The Large Area Telescope (LAT), which is a pair conversion
detector on board the \textsl{Fermi} Gamma-ray Space Telescope
\citep{Atwood2009}. \textsl{Fermi}/LAT has been in operation since 2008
August 11 and detects $\gamma$-rays with energies in the range from
$\sim$20\,MeV to $>$300\,GeV. The sensitivity of LAT is more than one
order of magnitude higher than that of EGRET. The whole sky is scanned by
LAT every 3 hours and in its sky-survey mode uniform exposure is obtained
within a few months. One of LATs major scientific goals is to observe the
$\gamma$-ray activity of AGN, i.e., the detection and monitoring and of
variable flaring sources. Most of the EGRET detections were confirmed. A
detailed discussion of AGN detections by \textsl{Fermi}/LAT is given by
\citet{3m_agn2009} based on the first three months of LAT observations.

Tracking Active Galactic Nuclei with Austral Milliarcsecond
Interferometry (TANAMI) is a radio VLBI monitoring program of AGN jets
south of $-30^\circ$ declination. The observations are performed
approximately every two months with the telescopes of the Australian Long
Baseline Array \citep[LBA;][]{Norris1988} in combination with telescopes
in South Africa, Antarctica and Chile. A detailed discussion of the
TANAMI program including a description of the initial source sample,
explanations of the observation and data reduction procedures, and brief
notes on the individual sources are given by Ojha et al.~(submitted to
A\&A) and presented in contributions to this volume by
\citet{Ojha2009_fs} and \citet{Mueller2009_fs}. Observations are made at
8.4\,GHz and 22\,GHz yielding spectral indices of parsec-scale jet
features. VLBI monitoring allows to determine intrinsic jet parameters,
such as the jet speed and the collimation angle, and makes the tracking
of components in the jet possible. For that reason VLBI is a crucial
complement to high energy astronomy to figure out the underlying emission
processes of $\gamma$-rays in the jet.

Recent results of the MOJAVE team \citep{Lister2009_mojave} have shown
that the $\gamma$-ray brightness of AGN as seen by LAT is correlated with
VLBI jet properties \citep{Lister2009_gamma_radio,Kovalev2009}. MOJAVE is
an extensive long-term monitoring program of AGN jets in the northern
sky, begun in 1994. It provides a multi-year database to study
correlations between radio and $\gamma$-ray emission of blazar jets.

In this work, we present the $\gamma$-ray properties of the TANAMI
sample, which were obtained in the first year of \textsl{Fermi}
operations. The combined analysis of radio and $\gamma$-ray properties is
still in progress.

\section{Analysis}

We analyzed the $\gamma$-ray emission of the sources in the TANAMI sample
consisting of 65 AGN as of 2009 November. The sample, which was recently extended to 75
sources, is discussed by \citet{Mueller2009_fs}. We searched for
associations with $\gamma$-ray sources from the 1-year catalog under
development by the LAT team and determined upper limits on the
$\gamma$-ray flux for TANAMI sources without a $\gamma$-ray counterpart
in this catalog.

To obtain a first tentative association of an AGN from the sample a
simple comparison of the positions of the TANAMI sources with that of the
$\gamma$-ray sources in the preliminary version of First LAT Catalog
(Abdo et al., in preparation) turned out to be an efficient technique. In
most cases the result was unique, i.e., for each AGN there were not more
than one $\gamma$-ray source with a position uncertainty larger than the
separation to the radio position of the AGN. The tentative associations
were then confirmed by taking the automatic source association pipeline
of the LAT team \citet{LBAS2009} into account.
This process is based on positional coincidence but additionally
considers the class and physical expecations of possible association
candidates for a $\gamma$-ray source in order to determine association
probabilities. The application of a figure-of-merit (FoM) approach for
finding AGN associations is discussed by \citet{Healey2008}.

For AGN which could not be associated with $\gamma$-ray catalog sources
we calculated upper limits on the $\gamma$-ray flux and list the test
statistic (TS) for a $\gamma$-ray source at this position using a maximum
likelihood analysis \citep{Cash1979,Mattox1996}. $\sqrt{\textnormal{TS}}$
is comparable to the significance in sigma. For the calculation of upper
limits we use only photons in the ``Diffuse'' class with energies in the
range 100\,MeV -- 100\,GeV. To minimize contamination from limb
$\gamma$-rays, photons with zenith angles greater than $105^\circ$ were
removed.  The standard \textsl{Fermi}-LAT {\it ScienceTools} software
package
\footnote{\url{http://fermi.gsfc.nasa.gov/ssc/data/analysis/documentation/Cicerone/}}
(version v9r15p2) was used with the ``P6\_V3\_DIFFUSE'' set of instrument
response functions. The flux, photon index, and test statistic of each
source were determined by analyzing a Region of Interest (RoI) of
12$^{\circ}$ in radius centered at the radio position and we modeled the
LAT point sources with a single power-law ($dN/dE = K(E/E_0)^{-\Gamma}$).
The Galactic diffuse background and the isotropic background used in the
RoI model (including the $\gamma$-ray diffuse and residual instrumental
backgrounds) are the currently recommended versions released and
described in more detail in documentation available at the \textsl{Fermi}
Science Support Center
\footnote{\url{http://fermi.gsfc.nasa.gov/ssc/data/access/lat/BackgroundModels.html}}.

In addition to the TS values obtained in our upper limit analysis we
present TS maps of the analyzed region and the pointlike TS values.
Localized peaks in these TS maps indicate tentative new detections of a
source. The method of the pointlike TS is described in detail in a
contribution to this volume by \citet{Burnett2009_fs}.

\section{Results}

The analysis of LAT data revealed different results for different classes
of AGN. The sources in our sample are classified according to the
V\'eron-V\'eron 12th edition catalog \citep{Veron2006}. The detection
rates in the $\gamma$-ray regime of certain source classes is biased by
our sample selection. The TANAMI sample consists of a flux-limited radio
subsample combined with a $\gamma$-ray selected subsample based on known
and candidate EGRET detections. In addition sources from special classes,
such as intra-day variable (IDV) and giga-hertz peaked spectrum (GPS)
sources were included in the sample. During the operation time of
\textsl{Fermi} we added southern AGN, which are candidate counterparts
for LAT sources, to the inital sample of 43 AGN.

\begin{figure}
    \includegraphics[width=0.9\columnwidth]{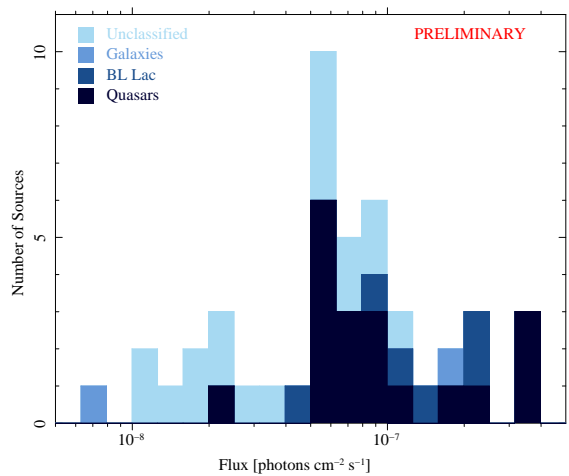}
    \caption{Flux distribution of the associated sources. The flux is
      given in units of photons\,cm$^{-2}$\,s$^{-1}$ in the energy range
      100\,MeV -- 100\,GeV.}
    \label{fig:fluxdistr}
\end{figure}
 
45 out of the 65 AGN from the TANAMI sample can be associated with
$\gamma$-ray sources from the 1-year catalog under development by the LAT
team. All of the 7 BL\,Lac objects in the sample were detected in the
$\gamma$-ray regime. 20 out of 28 quasars were detected, but only 2 out
of 11 radio galaxies have strong enough $\gamma$-ray emission to be
detected with LAT. The two detected radio galaxies are PKS\,0625$-$354
(OH$-$342) and PKS\,1322$-$428 (Cen\,A). This result is consistent with
the unification scheme \citep{Urry1995} in which the inclination of the jet
and thus the relativistic beaming effects are considered. Cen\,A is the
closest AGN and PKS\,0625$-$354 is interesting because it exhibits an
FR\,I radio-galaxy morphology whereas its optical spectrum is similar to
that of a BL\,Lac object \citep{Wills2004}. Further TANAMI observations
are necessary to determine properties of the inner radio jet of this
source, such as its speed. 

\begin{figure}
    \includegraphics[width=0.9\columnwidth]{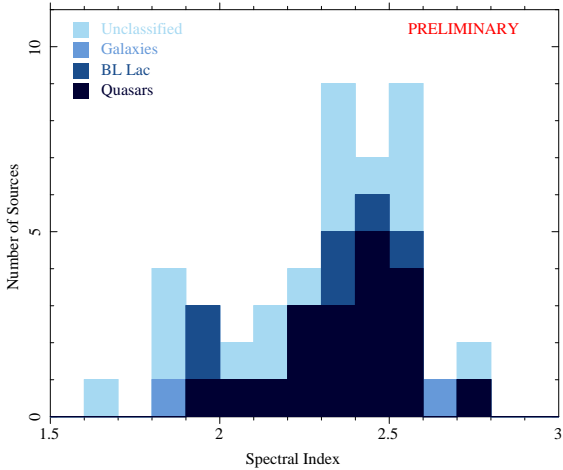}
    \caption{Distribution of the spectral indices $\Gamma$ of the
    associated sources.} \label{fig:indxdistr}
\end{figure}

The distribution of the $\gamma$-ray flux of the associated sources is shown in
Fig.~\ref{fig:fluxdistr}. The distribution of spectral indices
is presented in Fig.~\ref{fig:indxdistr}.

For sources that were not associated we performed a maximum likelihood
analysis to obtain upper limits on the flux. For this analysis we used
the positions of the AGN, which are know from observations at radio
frequencies, and analyzed the LAT data at these positions.

\begin{figure*}
    {\centering
    \includegraphics[width=.28\textwidth]{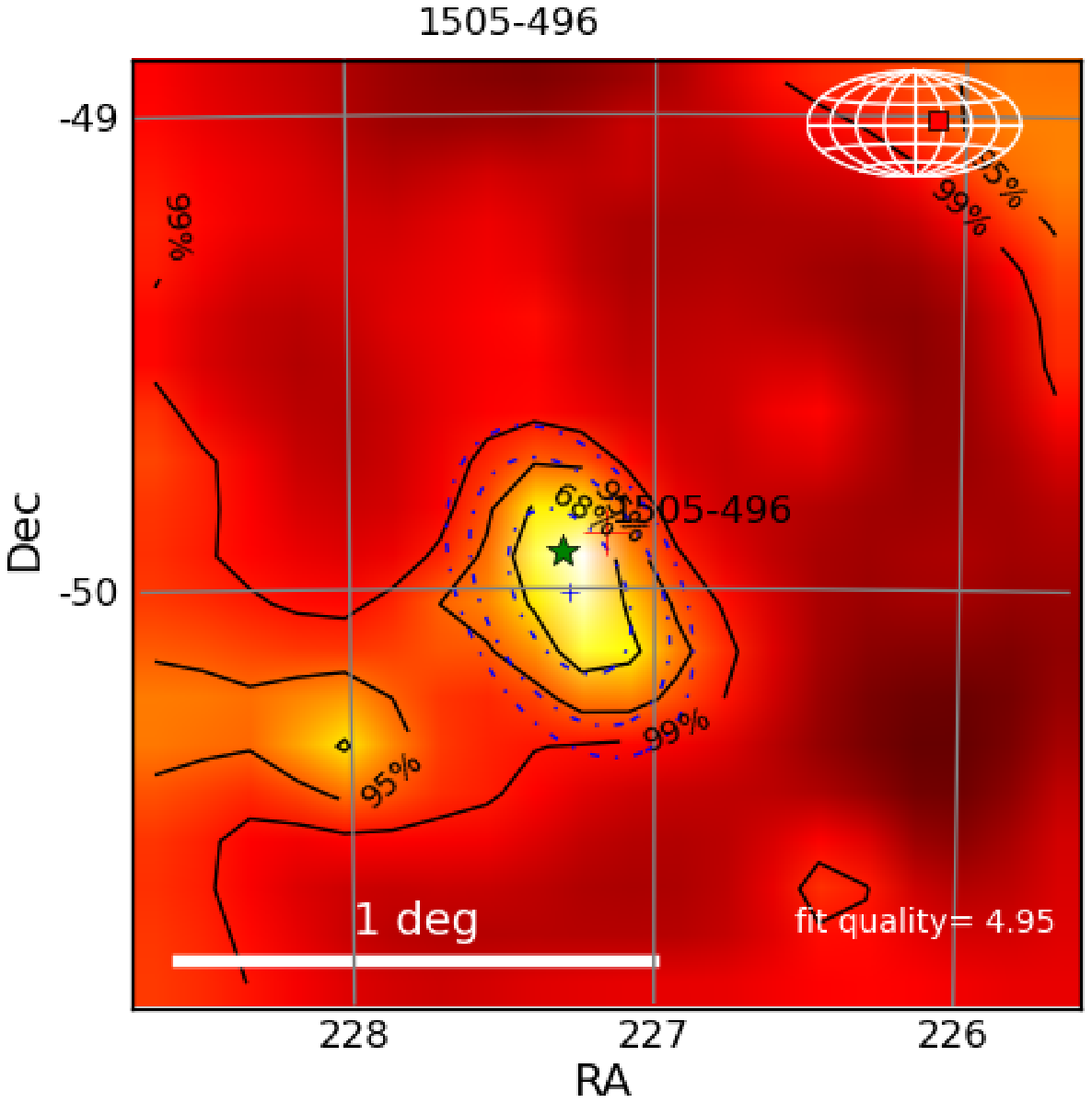}\hspace{25pt}
    \includegraphics[width=.28\textwidth]{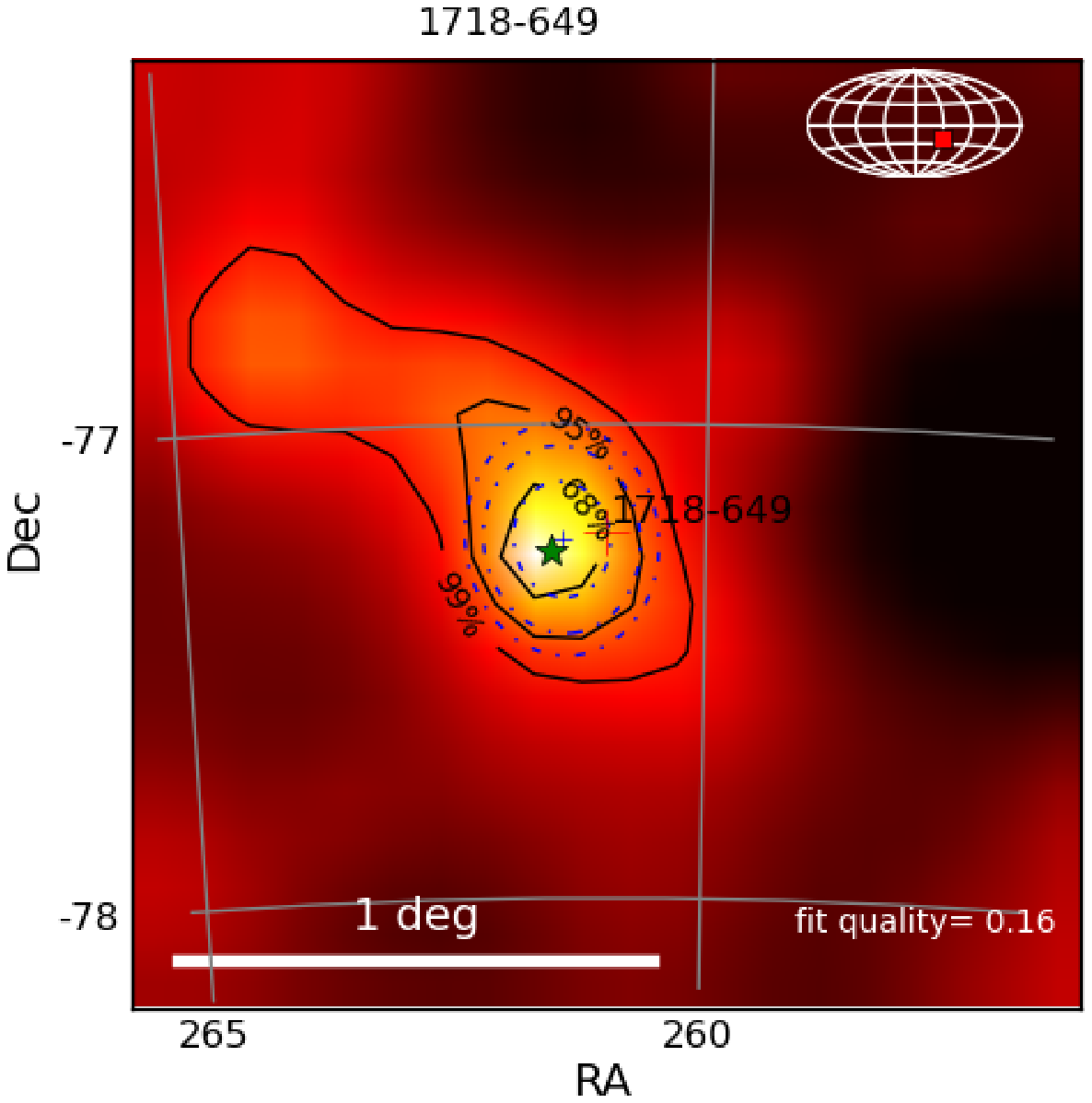}\\[5pt]
    \includegraphics[width=.28\textwidth]{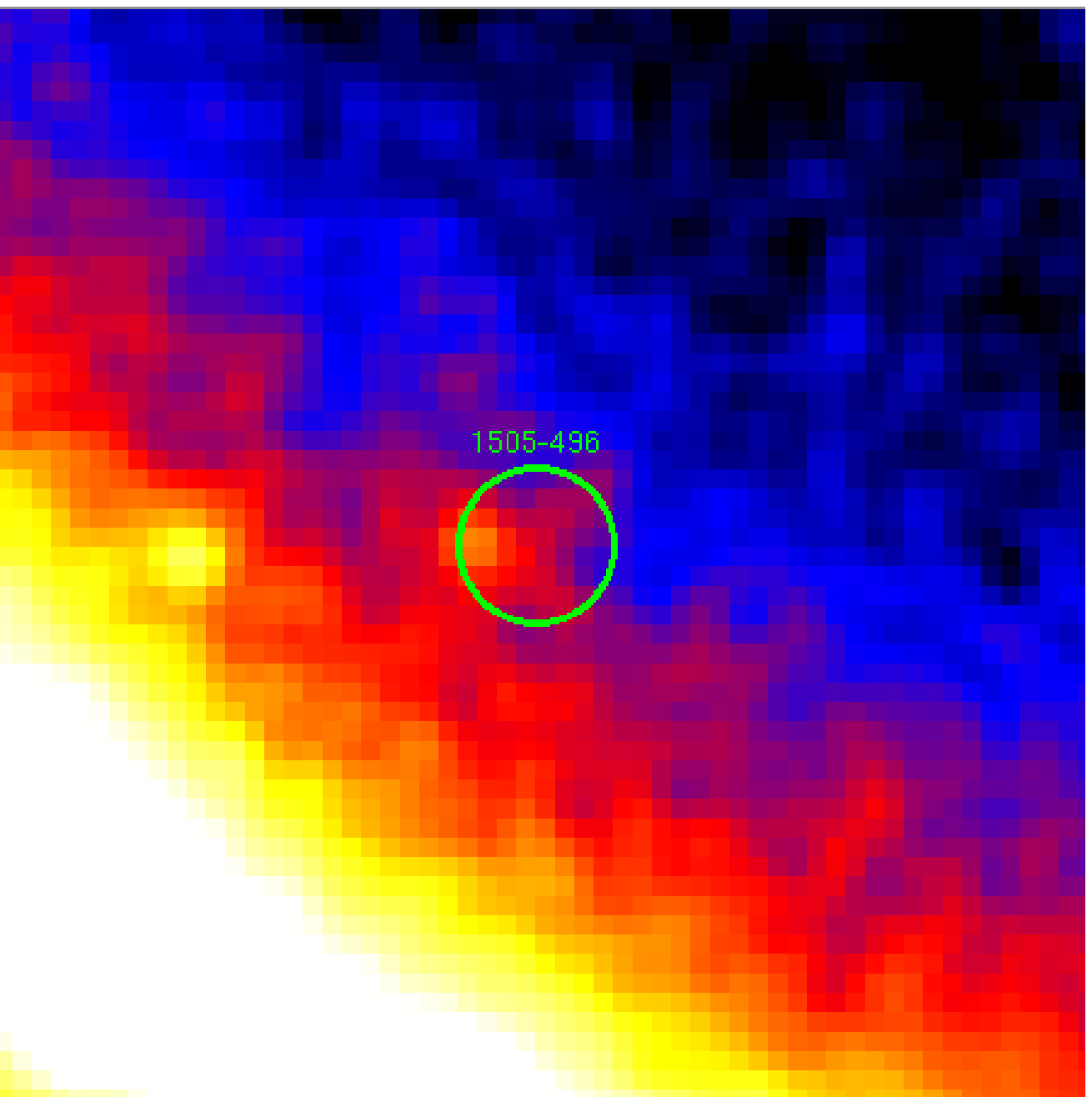}\hspace{25pt}
    \includegraphics[width=.28\textwidth]{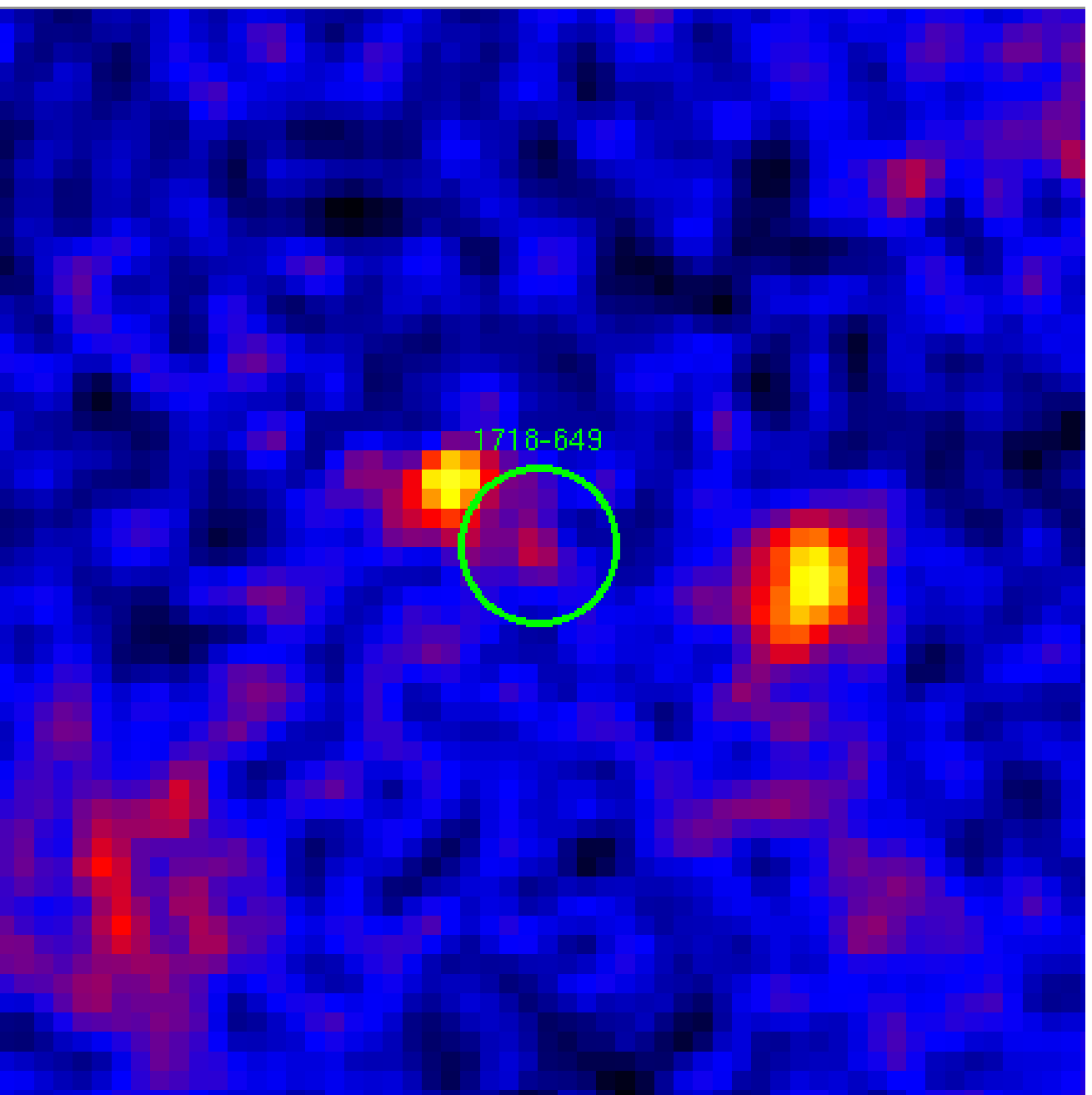}\\[20pt] 
    \includegraphics[width=.28\textwidth]{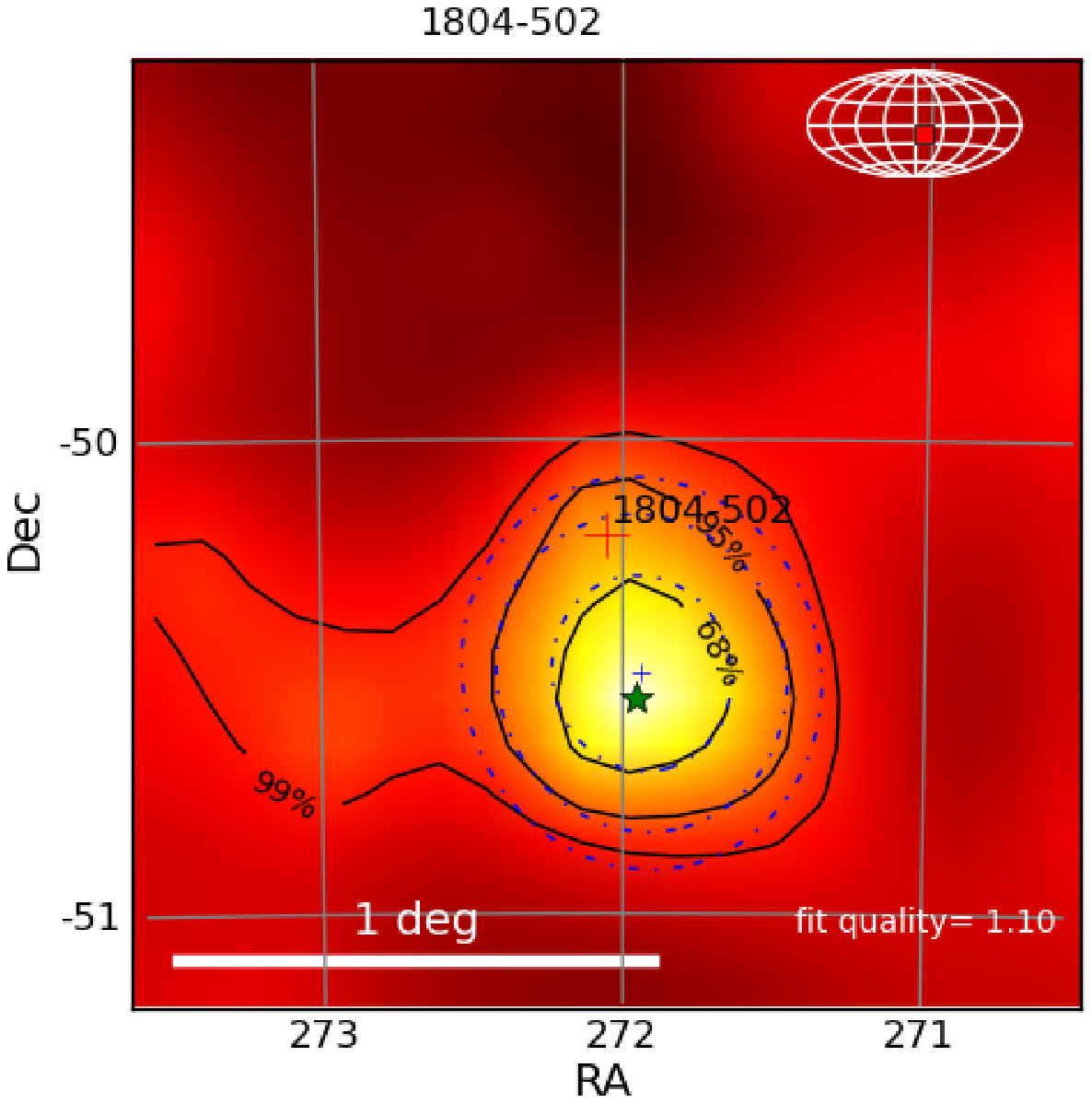}\hspace{25pt}
    \includegraphics[width=.28\textwidth]{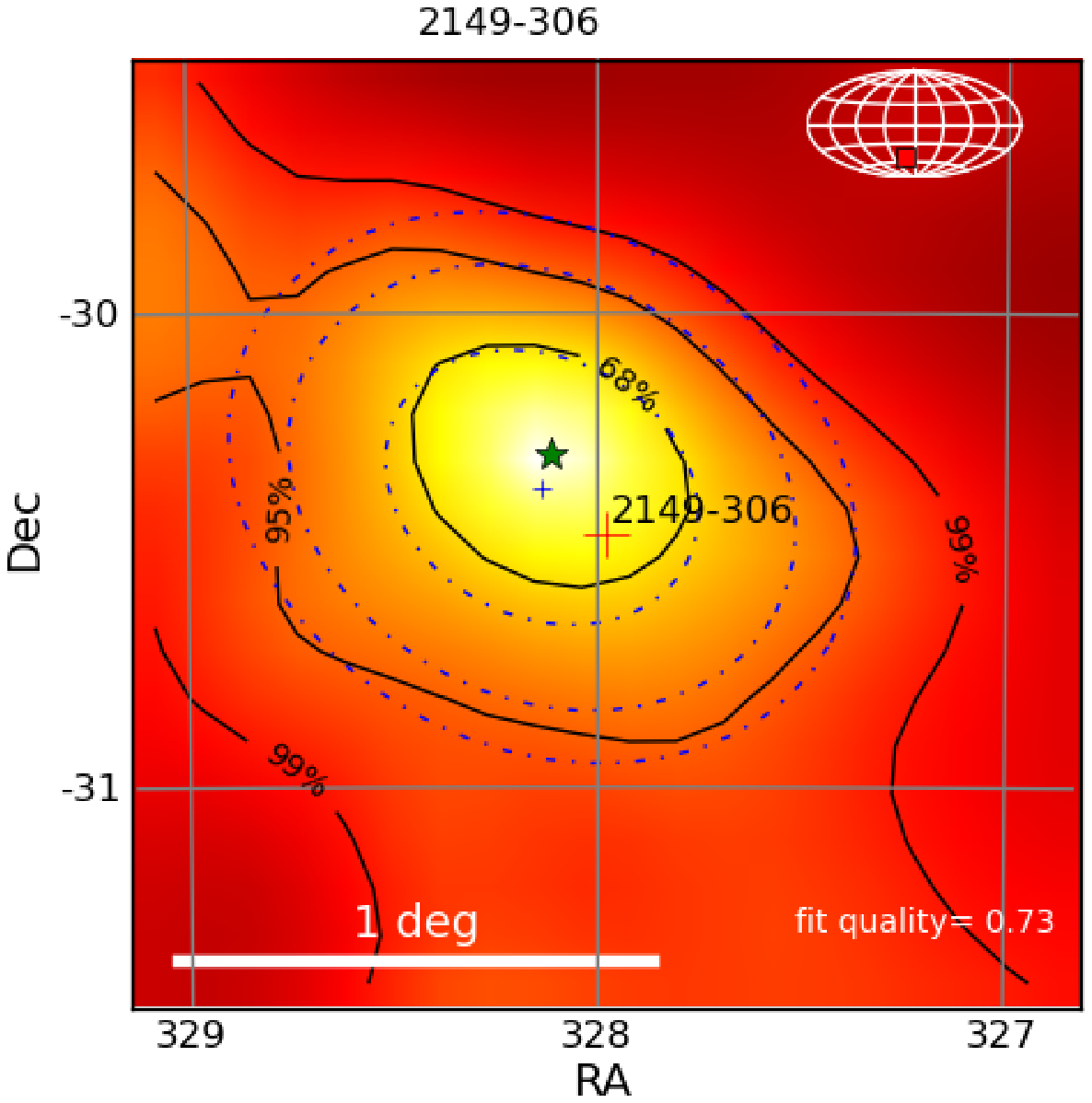}\hspace{25pt}
    \includegraphics[width=.28\textwidth]{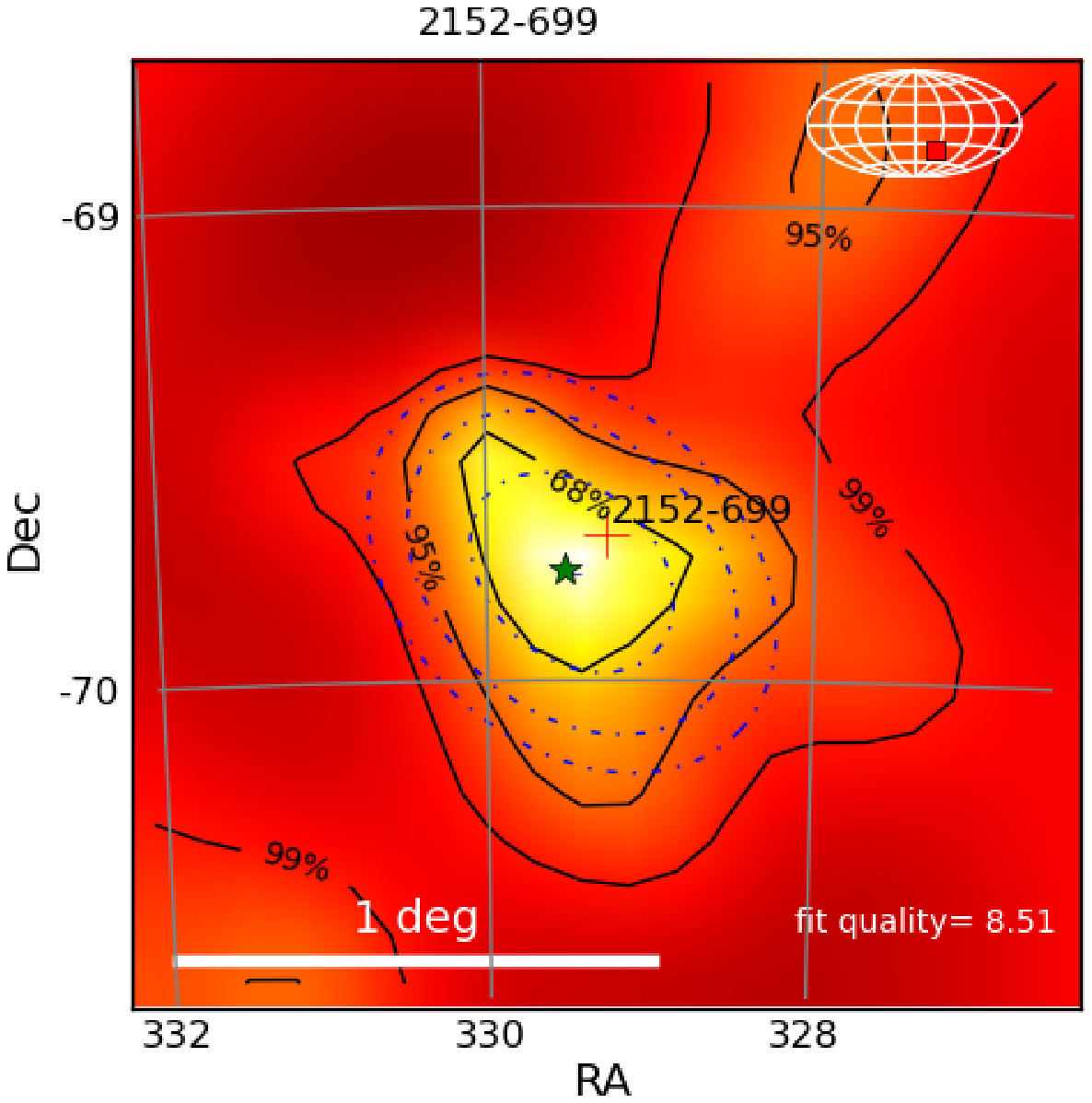}\\[5pt]
    \includegraphics[width=.28\textwidth]{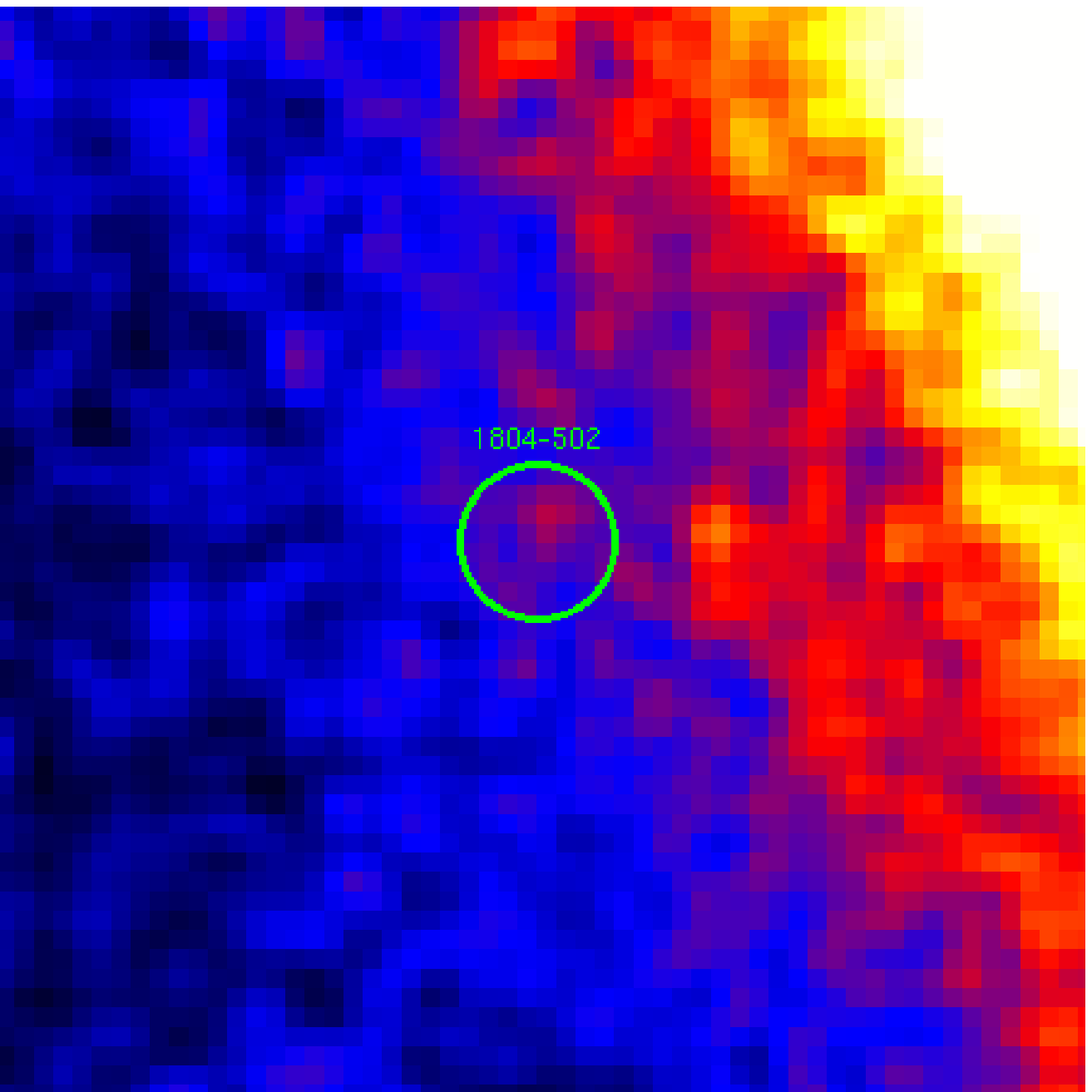}\hspace{25pt}
    \includegraphics[width=.28\textwidth]{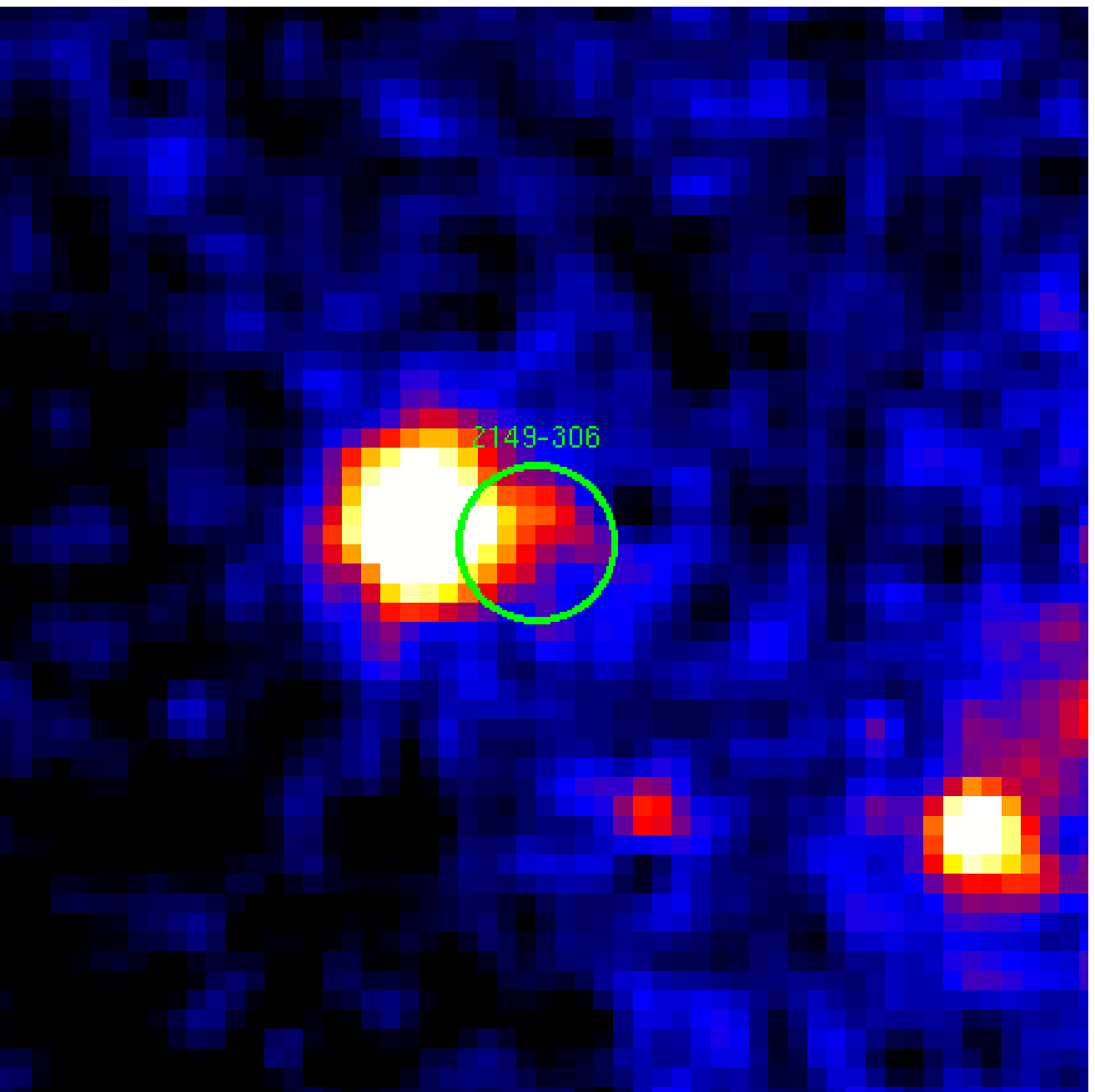}\hspace{25pt}
    \includegraphics[width=.28\textwidth]{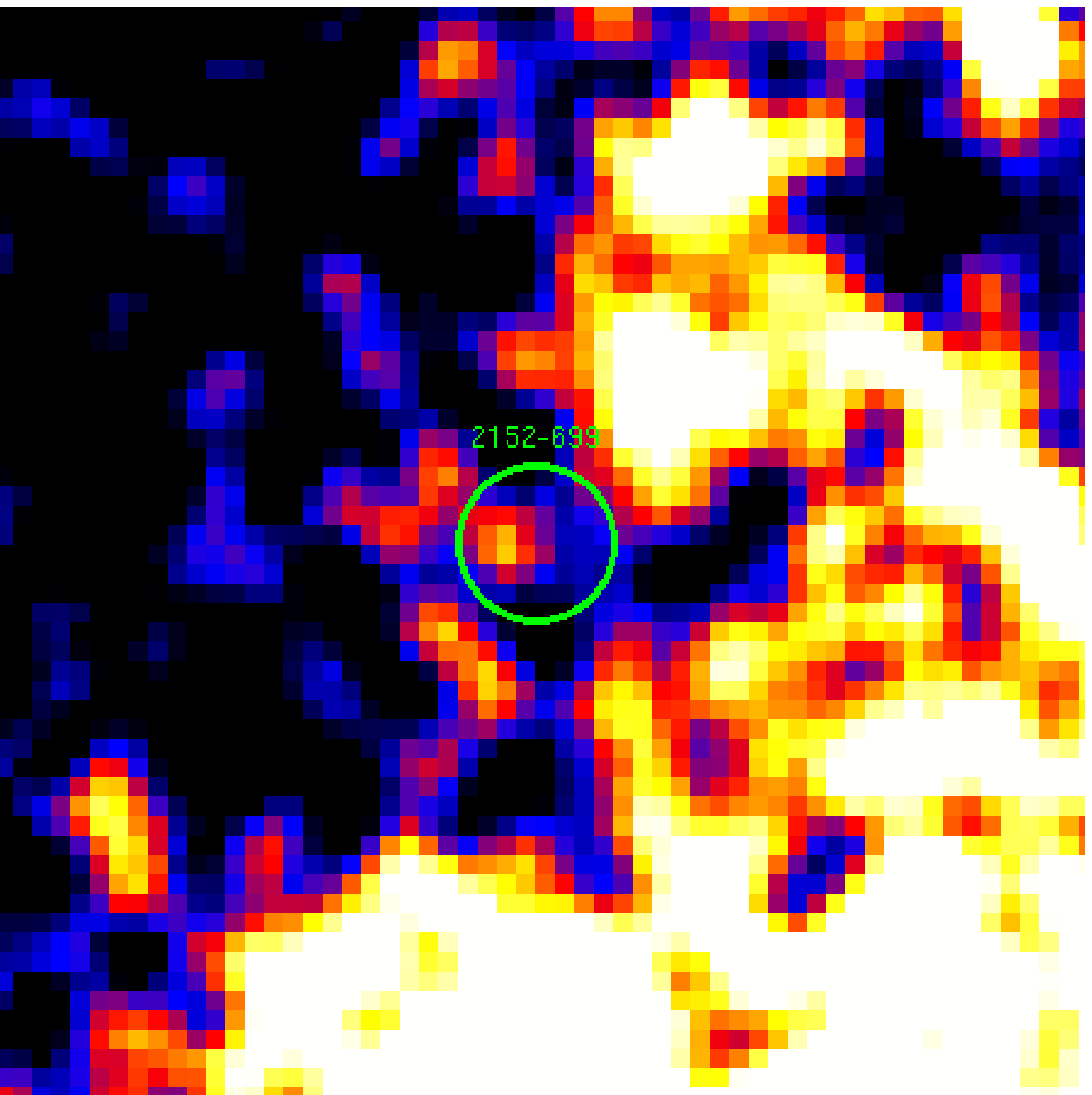}\\}
    \caption{TS maps, and below the corresponding count maps of possible
    new detections of five AGN (PKS\,1505$-$496, PKS\,1718$-$649,
    PKS\,1804$-$502, PKS\,2149$-$306, PKS\,2152$-$699). The diameter of
    the green circle in the count maps, which indicates the position of
    the analyzed source, is $2^\circ$.} \label{fig:maps}
\end{figure*}

The calculation of upper limits on the $\gamma$-ray flux for the
unassociated AGN from the TANAMI sample revealed in several cases a
relatively high TS. We modeled sources with $\textnormal{TS}>25$
following the criteria of the LAT team, while upper limits are presented
for the other sources. The resulting parameters are shown in
Tab.~\ref{tab:ts}. Out of these 20 sources we found two objects with test
statistic larger than 25. In both cases also the TS maps, which are
presented in Fig.~\ref{fig:maps}, show a localized peak at the
corresponding position.

The spectral indices of the tentative detections (Tab.~\ref{tab:ts}) are
larger than the average of the catalog sources shown in
Fig.~\ref{fig:indxdistr}. Modelling the other sources in the sample, with
$10<\textnormal{TS}<25$, yielded soft spectra as well. This trend can be
understood as the size of the point spread function of the LAT decreases
with energy, and makes it difficult to localize soft weak sources. These
tentative detections need to be confirmed by further analysis, such as
the comparison with a large sample of random positions. For that reason
only upper limits for these sources are presented here. Several of the
tentatively detected sources are close to bright sources or not far away
from the Galactic plane. Further analysis is required to determine if the
asymmetric distribution of spectral indices is a characteristic of the
sample.

In addition to the likelihood analysis, we calculated TS maps
\citep{Burnett2009_fs}. The pointlike TS peak values from the TS maps are
qualitatively similar but show a significant quantitative difference to
the TS values obtained with the upper limit analysis. This differences
will be investigated in more detail, such as the analysis of a large sample of
random position.

\begin{table}
    \caption{$\gamma$-ray flux and upper limits of the unassociated
    AGN from the sample.} \label{tab:ts}
    \begin{tabular}{rrrrrr}
	\hline \hline Name & ID\footnote{~The optical counterpart,
	denoted as follows: (Q) quasar, (B) BL Lac object, (G)
	galaxy, (U) unclassified} & TS & ~\quad\quad\quad
	Flux\footnote{~Flux in the energy band from 100\,MeV to
	100\,GeV in units of
	[$\times10^{-8}$\,\,photons\,\,cm$^{-2}$\,\,s$^{-1}$]} &
	$\Gamma$ & ptlikeTS\\
	\hline 1505-496 & U & 51.64 &
	$6.2\pm1.5$ & $2.61\pm0.13$ & 5.4\\
	2149-306 & Q & 31.00 & $4.2\pm0.9$ & $3.1\pm0.2$ & 5.7\\
	\hline
	0438-436 & Q &  0.01 & $<3.79$ & --- & 0.6\\
	0518-458 & G &  9.18 & $<4.0$  & --- & 6.6\\
	0527-359 & U &  0.00 & $<0.30$ & --- & 0.0\\
	1104-445 & Q &  3.65 & $<3.1$  & --- & 0.7\\
	1257-326 & Q &  0.00 & $<0.23$ & --- & 3.3\\
	1333-337 & G &  0.00 & $<0.14$ & --- & 0.0\\
	1549-790 & G &  2.04 & $<9.0$  & --- & 0.0\\
	1716-771 & U &  0.00 & $<1.44$ & --- & 1.9\\
	1718-649 & G & 14.00 & $<2.4$  & --- & 8.8\\
	1733-565 & G &  0.00 & $<1.73$ & --- & 0.0\\
	1804-502 & Q & 18.32 & $<8.0$  & --- & 7.0\\
	1814-637 & G &  1.56 & $<5.0$  & --- & 1.8\\
	1934-638 & G &  0.00 & $<0.15$ & --- & 0.0\\
	2027-308 & G & 14.29 & $<3.5$  & --- & 2.0\\
	2106-413 & Q &  0.00 & $<0.13$ & --- & 0.0\\
	2152-699 & G & 20.32 & $<5.8$  & --- & 6.7\\
	2326-477 & Q & 17.56 & $<3.7$  & --- & 1.1\\
	2355-534 & Q & 15.03 & $<8.0$  & --- & 1.6\\
	\hline
    \end{tabular}
\end{table}

In Fig.~\ref{fig:maps}, the TS maps of the five unassociated sources with
the best localized TS peak, and the corresponding count maps are shown.
The source PKS\,1718$-$649 is an interesting candidate, as this source is
one of the closest (56\,Mpc) and best studied GPS sources
\citep[e.g.,][]{Tingay2003b} and would be the first LAT detected source
of this type. The radio morphology of this source is classified as
irregular as it does not fit in the classification scheme of ``compact'',
``single-sided jet'', or ``double sided jet''. More TANAMI observations
of this source have to be analyzed to clarify its structure and
especially the kinematic behavior of the radio jet.

\section{Summary}

We analyzed the $\gamma$-ray properties of the TANAMI sample based on the
data obtained with \textsl{Fermi}-LAT during its first year of operation.
45 out of the 65 AGN from this sample can be associated with $\gamma$-ray
sources from the 1-year catalog under development by the LAT team. All
BL\,Lac objects and a large fraction of quasars (71\%) are detected in
the $\gamma$-ray regime, whereas from the 11 radio galaxies only the
closest one (Cen\,A) and one (PKS\,0625$-$354) with an optical spectrum
similar to a BL\,Lac object were detected with \textsl{Fermi}-LAT. For
the 20 AGN without $\gamma$-ray counterpart we presented upper limits on
the $\gamma$-ray flux. In two cases (PKS\,1505$-$496, PKS\,2149$-$306)
the TS was high enough to model a source. The combination with TS maps
indicated tentative new $\gamma$-ray detections of five sources
(PKS\,1505$-$496, PKS\,1718$-$649, PKS\,1804$-$502, PKS\,2149$-$306,
PKS\,2152$-$699).

\bigskip
\begin{acknowledgments}
    This research has been partially funded by the Bundesministerium f\"ur
    Wirtschaft und Technologie under Deutsches Zentrum f\"ur Luft- und
    Raumfahrt grant number 50OR0808.
    The Long Baseline Array is part of the Australia Telescope which is
    funded by the Commonwealth of Australia for operation as a National
    Facility managed by CSIRO.
\end{acknowledgments}


\bigskip

\begin{thebibliography}{23}
\expandafter\ifx\csname natexlab\endcsname\relax\def\natexlab#1{#1}\fi
\expandafter\ifx\csname bibnamefont\endcsname\relax
  \def\bibnamefont#1{#1}\fi
\expandafter\ifx\csname bibfnamefont\endcsname\relax
  \def\bibfnamefont#1{#1}\fi
\expandafter\ifx\csname citenamefont\endcsname\relax
  \def\citenamefont#1{#1}\fi
\expandafter\ifx\csname url\endcsname\relax
  \def\url#1{\texttt{#1}}\fi
\expandafter\ifx\csname urlprefix\endcsname\relax\def\urlprefix{URL }\fi
\providecommand{\bibinfo}[2]{#2}
\providecommand{\eprint}[2][]{\url{#2}}

\bibitem[{\citenamefont{{Blandford} and {Rees}}(1978)}]{Blandford1978}
\bibinfo{author}{\bibfnamefont{R.~D.} \bibnamefont{{Blandford}}}
  \bibnamefont{and} \bibinfo{author}{\bibfnamefont{M.~J.}
  \bibnamefont{{Rees}}}, in \emph{\bibinfo{booktitle}{BL Lac Objects}}, edited
  by \bibinfo{editor}{\bibnamefont{{A.~M.~Wolfe}}} (\bibinfo{year}{1978}), pp.
  \bibinfo{pages}{328}.

\bibitem[{\citenamefont{{Maraschi} et~al.}(1992)\citenamefont{{Maraschi},
  {Ghisellini}, and {Celotti}}}]{Maraschi1992}
\bibinfo{author}{\bibfnamefont{L.}~\bibnamefont{{Maraschi}}},
  \bibinfo{author}{\bibfnamefont{G.}~\bibnamefont{{Ghisellini}}},
  \bibnamefont{and}
  \bibinfo{author}{\bibfnamefont{A.}~\bibnamefont{{Celotti}}},
  \bibinfo{journal}{ApJL} \textbf{\bibinfo{volume}{397}}, \bibinfo{pages}{L5}
  (\bibinfo{year}{1992}).

\bibitem[{\citenamefont{{Lister}
  et~al.}(2009{\natexlab{a}})\citenamefont{{Lister}, {Cohen}, {Homan},
  {Kadler}, {Kellermann}, {Kovalev}, {Ros}, {Savolainen}, and
  {Zensus}}}]{Lister2009_kinematics}
\bibinfo{author}{\bibfnamefont{M.~L.} \bibnamefont{{Lister}}},
  \bibinfo{author}{\bibfnamefont{M.~H.} \bibnamefont{{Cohen}}},
  \bibinfo{author}{\bibfnamefont{D.~C.} \bibnamefont{{Homan}}},
  \bibinfo{author}{\bibfnamefont{M.}~\bibnamefont{{Kadler}}},
  \bibinfo{author}{\bibfnamefont{K.~I.} \bibnamefont{{Kellermann}}},
  \bibinfo{author}{\bibfnamefont{Y.~Y.} \bibnamefont{{Kovalev}}},
  \bibinfo{author}{\bibfnamefont{E.}~\bibnamefont{{Ros}}},
  \bibinfo{author}{\bibfnamefont{T.}~\bibnamefont{{Savolainen}}},
  \bibnamefont{and} \bibinfo{author}{\bibfnamefont{J.~A.}
  \bibnamefont{{Zensus}}}, \bibinfo{journal}{AJ}
  \textbf{\bibinfo{volume}{138}}, \bibinfo{pages}{1874}
  (\bibinfo{year}{2009}{\natexlab{a}}).

\bibitem[{\citenamefont{{Thompson} et~al.}(1993)\citenamefont{{Thompson},
  {Bertsch}, {Fichtel}, {Hartman}, {Hofstadter}, {Hughes}, {Hunter},
  {Hughlock}, {Kanbach}, {Kniffen} et~al.}}]{Thompson1993_egret}
\bibinfo{author}{\bibfnamefont{D.~J.} \bibnamefont{{Thompson}}},
  \bibinfo{author}{\bibfnamefont{D.~L.} \bibnamefont{{Bertsch}}},
  \bibinfo{author}{\bibfnamefont{C.~E.} \bibnamefont{{Fichtel}}},
  \bibinfo{author}{\bibfnamefont{R.~C.} \bibnamefont{{Hartman}}},
  \bibinfo{author}{\bibfnamefont{R.}~\bibnamefont{{Hofstadter}}},
  \bibinfo{author}{\bibfnamefont{E.~B.} \bibnamefont{{Hughes}}},
  \bibinfo{author}{\bibfnamefont{S.~D.} \bibnamefont{{Hunter}}},
  \bibinfo{author}{\bibfnamefont{B.~W.} \bibnamefont{{Hughlock}}},
  \bibinfo{author}{\bibfnamefont{G.}~\bibnamefont{{Kanbach}}},
  \bibinfo{author}{\bibfnamefont{D.~A.} \bibnamefont{{Kniffen}}},
  \bibnamefont{et~al.}, \bibinfo{journal}{ApJS} \textbf{\bibinfo{volume}{86}},
  \bibinfo{pages}{629} (\bibinfo{year}{1993}).

\bibitem[{\citenamefont{{Atwood} et~al.}(2009)\citenamefont{{Atwood}, {Abdo},
  {Ackermann}, {Althouse}, {Anderson}, {Axelsson}, {Baldini}, {Ballet}, {Band},
  {Barbiellini} et~al.}}]{Atwood2009}
\bibinfo{author}{\bibfnamefont{W.~B.} \bibnamefont{{Atwood}}},
  \bibinfo{author}{\bibfnamefont{A.~A.} \bibnamefont{{Abdo}}},
  \bibinfo{author}{\bibfnamefont{M.}~\bibnamefont{{Ackermann}}},
  \bibinfo{author}{\bibfnamefont{W.}~\bibnamefont{{Althouse}}},
  \bibinfo{author}{\bibfnamefont{B.}~\bibnamefont{{Anderson}}},
  \bibinfo{author}{\bibfnamefont{M.}~\bibnamefont{{Axelsson}}},
  \bibinfo{author}{\bibfnamefont{L.}~\bibnamefont{{Baldini}}},
  \bibinfo{author}{\bibfnamefont{J.}~\bibnamefont{{Ballet}}},
  \bibinfo{author}{\bibfnamefont{D.~L.} \bibnamefont{{Band}}},
  \bibinfo{author}{\bibfnamefont{G.}~\bibnamefont{{Barbiellini}}},
  \bibnamefont{et~al.}, \bibinfo{journal}{ApJ} \textbf{\bibinfo{volume}{697}},
  \bibinfo{pages}{1071} (\bibinfo{year}{2009}).

\bibitem[{\citenamefont{{Abdo} et~al.}(2009{\natexlab{a}})\citenamefont{{Abdo},
  {Ackermann}, {Ajello}, {Atwood}, {Axelsson}, {Baldini}, {Ballet},
  {Barbiellini}, {Bastieri}, {Baughman} et~al.}}]{3m_agn2009}
\bibinfo{author}{\bibfnamefont{A.~A.} \bibnamefont{{Abdo}}},
  \bibinfo{author}{\bibfnamefont{M.}~\bibnamefont{{Ackermann}}},
  \bibinfo{author}{\bibfnamefont{M.}~\bibnamefont{{Ajello}}},
  \bibinfo{author}{\bibfnamefont{W.~B.} \bibnamefont{{Atwood}}},
  \bibinfo{author}{\bibfnamefont{M.}~\bibnamefont{{Axelsson}}},
  \bibinfo{author}{\bibfnamefont{L.}~\bibnamefont{{Baldini}}},
  \bibinfo{author}{\bibfnamefont{J.}~\bibnamefont{{Ballet}}},
  \bibinfo{author}{\bibfnamefont{G.}~\bibnamefont{{Barbiellini}}},
  \bibinfo{author}{\bibfnamefont{D.}~\bibnamefont{{Bastieri}}},
  \bibinfo{author}{\bibfnamefont{B.~M.} \bibnamefont{{Baughman}}},
  \bibnamefont{et~al.}, \bibinfo{journal}{ApJ} \textbf{\bibinfo{volume}{700}},
  \bibinfo{pages}{597} (\bibinfo{year}{2009}{\natexlab{a}}).

\bibitem[{\citenamefont{{Norris}}(1988)}]{Norris1988}
\bibinfo{author}{\bibfnamefont{R.~P.} \bibnamefont{{Norris}}}, in
  \emph{\bibinfo{booktitle}{The Impact of VLBI on Astrophysics and
  Geophysics}}, edited by \bibinfo{editor}{\bibnamefont{{M.~J.~Reid \&
  J.~M.~Moran}}} (\bibinfo{year}{1988}),  \emph{\bibinfo{series}{IAU Symposium}}
  \bibinfo{volume}{129}, \bibinfo{pages}{485}.

\bibitem[{\citenamefont{{Ojha}~et al.}(2009)}]{Ojha2009_fs}
\bibinfo{author}{\bibfnamefont{R.}~\bibnamefont{{Ojha}~et al.}},
  \bibinfo{journal}{Proc. Fermi Symposium (this volume)}
  (\bibinfo{year}{2009}).

\bibitem[{\citenamefont{{M{\"u}ller}~et al.}(2009)}]{Mueller2009_fs}
\bibinfo{author}{\bibfnamefont{C.}~\bibnamefont{{M{\"u}ller}~et al.}},
  \bibinfo{journal}{Proc. Fermi Symposium (this volume)}
  (\bibinfo{year}{2009}).

\bibitem[{\citenamefont{{Lister}
  et~al.}(2009{\natexlab{b}})\citenamefont{{Lister}, {Aller}, {Aller}, {Cohen},
  {Homan}, {Kadler}, {Kellermann}, {Kovalev}, {Ros}, {Savolainen}
  et~al.}}]{Lister2009_mojave}
\bibinfo{author}{\bibfnamefont{M.~L.} \bibnamefont{{Lister}}},
  \bibinfo{author}{\bibfnamefont{H.~D.} \bibnamefont{{Aller}}},
  \bibinfo{author}{\bibfnamefont{M.~F.} \bibnamefont{{Aller}}},
  \bibinfo{author}{\bibfnamefont{M.~H.} \bibnamefont{{Cohen}}},
  \bibinfo{author}{\bibfnamefont{D.~C.} \bibnamefont{{Homan}}},
  \bibinfo{author}{\bibfnamefont{M.}~\bibnamefont{{Kadler}}},
  \bibinfo{author}{\bibfnamefont{K.~I.} \bibnamefont{{Kellermann}}},
  \bibinfo{author}{\bibfnamefont{Y.~Y.} \bibnamefont{{Kovalev}}},
  \bibinfo{author}{\bibfnamefont{E.}~\bibnamefont{{Ros}}},
  \bibinfo{author}{\bibfnamefont{T.}~\bibnamefont{{Savolainen}}},
  \bibnamefont{et~al.}, \bibinfo{journal}{AJ} \textbf{\bibinfo{volume}{137}},
  \bibinfo{pages}{3718} (\bibinfo{year}{2009}{\natexlab{b}}).

\bibitem[{\citenamefont{{Lister}
  et~al.}(2009{\natexlab{c}})\citenamefont{{Lister}, {Homan}, {Kadler},
  {Kellermann}, {Kovalev}, {Ros}, {Savolainen}, and
  {Zensus}}}]{Lister2009_gamma_radio}
\bibinfo{author}{\bibfnamefont{M.~L.} \bibnamefont{{Lister}}},
  \bibinfo{author}{\bibfnamefont{D.~C.} \bibnamefont{{Homan}}},
  \bibinfo{author}{\bibfnamefont{M.}~\bibnamefont{{Kadler}}},
  \bibinfo{author}{\bibfnamefont{K.~I.} \bibnamefont{{Kellermann}}},
  \bibinfo{author}{\bibfnamefont{Y.~Y.} \bibnamefont{{Kovalev}}},
  \bibinfo{author}{\bibfnamefont{E.}~\bibnamefont{{Ros}}},
  \bibinfo{author}{\bibfnamefont{T.}~\bibnamefont{{Savolainen}}},
  \bibnamefont{and} \bibinfo{author}{\bibfnamefont{J.~A.}
  \bibnamefont{{Zensus}}}, \bibinfo{journal}{ApJL}
  \textbf{\bibinfo{volume}{696}}, \bibinfo{pages}{L22}
  (\bibinfo{year}{2009}{\natexlab{c}}).

\bibitem[{\citenamefont{{Kovalev} et~al.}(2009)\citenamefont{{Kovalev},
  {Aller}, {Aller}, {Homan}, {Kadler}, {Kellermann}, {Kovalev}, {Lister},
  {McCormick}, {Pushkarev} et~al.}}]{Kovalev2009}
\bibinfo{author}{\bibfnamefont{Y.~Y.} \bibnamefont{{Kovalev}}},
  \bibinfo{author}{\bibfnamefont{H.~D.} \bibnamefont{{Aller}}},
  \bibinfo{author}{\bibfnamefont{M.~F.} \bibnamefont{{Aller}}},
  \bibinfo{author}{\bibfnamefont{D.~C.} \bibnamefont{{Homan}}},
  \bibinfo{author}{\bibfnamefont{M.}~\bibnamefont{{Kadler}}},
  \bibinfo{author}{\bibfnamefont{K.~I.} \bibnamefont{{Kellermann}}},
  \bibinfo{author}{\bibfnamefont{Y.~A.} \bibnamefont{{Kovalev}}},
  \bibinfo{author}{\bibfnamefont{M.~L.} \bibnamefont{{Lister}}},
  \bibinfo{author}{\bibfnamefont{M.~J.} \bibnamefont{{McCormick}}},
  \bibinfo{author}{\bibfnamefont{A.~B.} \bibnamefont{{Pushkarev}}},
  \bibnamefont{et~al.}, \bibinfo{journal}{ApJL} \textbf{\bibinfo{volume}{696}},
  \bibinfo{pages}{L17} (\bibinfo{year}{2009}).

\bibitem[{\citenamefont{{Abdo} et~al.}(2009{\natexlab{b}})\citenamefont{{Abdo},
  {Ackermann}, {Ajello}, {Atwood}, {Axelsson}, {Baldini}, {Ballet}, {Band},
  {Barbiellini}, {Bastieri} et~al.}}]{LBAS2009}
\bibinfo{author}{\bibfnamefont{A.~A.} \bibnamefont{{Abdo}}},
  \bibinfo{author}{\bibfnamefont{M.}~\bibnamefont{{Ackermann}}},
  \bibinfo{author}{\bibfnamefont{M.}~\bibnamefont{{Ajello}}},
  \bibinfo{author}{\bibfnamefont{W.~B.} \bibnamefont{{Atwood}}},
  \bibinfo{author}{\bibfnamefont{M.}~\bibnamefont{{Axelsson}}},
  \bibinfo{author}{\bibfnamefont{L.}~\bibnamefont{{Baldini}}},
  \bibinfo{author}{\bibfnamefont{J.}~\bibnamefont{{Ballet}}},
  \bibinfo{author}{\bibfnamefont{D.~L.} \bibnamefont{{Band}}},
  \bibinfo{author}{\bibfnamefont{G.}~\bibnamefont{{Barbiellini}}},
  \bibinfo{author}{\bibfnamefont{D.}~\bibnamefont{{Bastieri}}},
  \bibnamefont{et~al.}, \bibinfo{journal}{ApJS} \textbf{\bibinfo{volume}{183}},
  \bibinfo{pages}{46} (\bibinfo{year}{2009}{\natexlab{b}}).

\bibitem[{\citenamefont{{Healey} et~al.}(2008)\citenamefont{{Healey}, {Romani},
  {Cotter}, {Michelson}, {Schlafly}, {Readhead}, {Giommi}, {Chaty}, {Grenier},
  and {Weintraub}}}]{Healey2008}
\bibinfo{author}{\bibfnamefont{S.~E.} \bibnamefont{{Healey}}},
  \bibinfo{author}{\bibfnamefont{R.~W.} \bibnamefont{{Romani}}},
  \bibinfo{author}{\bibfnamefont{G.}~\bibnamefont{{Cotter}}},
  \bibinfo{author}{\bibfnamefont{P.~F.} \bibnamefont{{Michelson}}},
  \bibinfo{author}{\bibfnamefont{E.~F.} \bibnamefont{{Schlafly}}},
  \bibinfo{author}{\bibfnamefont{A.~C.~S.} \bibnamefont{{Readhead}}},
  \bibinfo{author}{\bibfnamefont{P.}~\bibnamefont{{Giommi}}},
  \bibinfo{author}{\bibfnamefont{S.}~\bibnamefont{{Chaty}}},
  \bibinfo{author}{\bibfnamefont{I.~A.} \bibnamefont{{Grenier}}},
  \bibnamefont{and} \bibinfo{author}{\bibfnamefont{L.~C.}
  \bibnamefont{{Weintraub}}}, \bibinfo{journal}{ApJS}
  \textbf{\bibinfo{volume}{175}}, \bibinfo{pages}{97} (\bibinfo{year}{2008}).

\bibitem[{\citenamefont{{Cash}}(1979)}]{Cash1979}
\bibinfo{author}{\bibfnamefont{W.}~\bibnamefont{{Cash}}},
  \bibinfo{journal}{ApJ} \textbf{\bibinfo{volume}{228}}, \bibinfo{pages}{939}
  (\bibinfo{year}{1979}).

\bibitem[{\citenamefont{{Mattox} et~al.}(1996)\citenamefont{{Mattox},
  {Bertsch}, {Chiang}, {Dingus}, {Digel}, {Esposito}, {Fierro}, {Hartman},
  {Hunter}, {Kanbach} et~al.}}]{Mattox1996}
\bibinfo{author}{\bibfnamefont{J.~R.} \bibnamefont{{Mattox}}},
  \bibinfo{author}{\bibfnamefont{D.~L.} \bibnamefont{{Bertsch}}},
  \bibinfo{author}{\bibfnamefont{J.}~\bibnamefont{{Chiang}}},
  \bibinfo{author}{\bibfnamefont{B.~L.} \bibnamefont{{Dingus}}},
  \bibinfo{author}{\bibfnamefont{S.~W.} \bibnamefont{{Digel}}},
  \bibinfo{author}{\bibfnamefont{J.~A.} \bibnamefont{{Esposito}}},
  \bibinfo{author}{\bibfnamefont{J.~M.} \bibnamefont{{Fierro}}},
  \bibinfo{author}{\bibfnamefont{R.~C.} \bibnamefont{{Hartman}}},
  \bibinfo{author}{\bibfnamefont{S.~D.} \bibnamefont{{Hunter}}},
  \bibinfo{author}{\bibfnamefont{G.}~\bibnamefont{{Kanbach}}},
  \bibnamefont{et~al.}, \bibinfo{journal}{ApJ} \textbf{\bibinfo{volume}{461}},
  \bibinfo{pages}{396} (\bibinfo{year}{1996}).

\bibitem[{Not({\natexlab{a}})}]{Note1}
\bibinfo{note}{\protect \burl
  {http://fermi.gsfc.nasa.gov/ssc/data/analysis/documentation/Cicerone/}}.

\bibitem[{Not({\natexlab{b}})}]{Note2}
\bibinfo{note}{\protect \burl
  {http://fermi.gsfc.nasa.gov/ssc/data/access/lat/BackgroundModels.html}}.

\bibitem[{\citenamefont{{Burnett}}(2009)}]{Burnett2009_fs}
\bibinfo{author}{\bibfnamefont{T.}~\bibnamefont{{Burnett}}},
  \bibinfo{journal}{Proc. Fermi Symposium (this volume)}
  (\bibinfo{year}{2009}).

\bibitem[{\citenamefont{{V{\'e}ron-Cetty} and {V{\'e}ron}}(2006)}]{Veron2006}
\bibinfo{author}{\bibfnamefont{M.}~\bibnamefont{{V{\'e}ron-Cetty}}}
  \bibnamefont{and}
  \bibinfo{author}{\bibfnamefont{P.}~\bibnamefont{{V{\'e}ron}}},
  \bibinfo{journal}{A\&A} \textbf{\bibinfo{volume}{455}}, \bibinfo{pages}{773}
  (\bibinfo{year}{2006}).

\bibitem[{\citenamefont{{Urry} and {Padovani}}(1995)}]{Urry1995}
\bibinfo{author}{\bibfnamefont{C.~M.} \bibnamefont{{Urry}}} \bibnamefont{and}
  \bibinfo{author}{\bibfnamefont{P.}~\bibnamefont{{Padovani}}},
  \bibinfo{journal}{PASP} \textbf{\bibinfo{volume}{107}}, \bibinfo{pages}{803}
  (\bibinfo{year}{1995}).

\bibitem[{\citenamefont{{Wills} et~al.}(2004)\citenamefont{{Wills}, {Morganti},
  {Tadhunter}, {Robinson}, and {Villar-Martin}}}]{Wills2004}
\bibinfo{author}{\bibfnamefont{K.~A.} \bibnamefont{{Wills}}},
  \bibinfo{author}{\bibfnamefont{R.}~\bibnamefont{{Morganti}}},
  \bibinfo{author}{\bibfnamefont{C.~N.} \bibnamefont{{Tadhunter}}},
  \bibinfo{author}{\bibfnamefont{T.~G.} \bibnamefont{{Robinson}}},
  \bibnamefont{and}
  \bibinfo{author}{\bibfnamefont{M.}~\bibnamefont{{Villar-Martin}}},
  \bibinfo{journal}{MNRAS} \textbf{\bibinfo{volume}{347}}, \bibinfo{pages}{771}
  (\bibinfo{year}{2004}).

\bibitem[{\citenamefont{{Tingay} et~al.}(2003)\citenamefont{{Tingay},
  {Edwards}, and {Tzioumis}}}]{Tingay2003b}
\bibinfo{author}{\bibfnamefont{S.~J.} \bibnamefont{{Tingay}}},
  \bibinfo{author}{\bibfnamefont{P.~G.} \bibnamefont{{Edwards}}},
  \bibnamefont{and} \bibinfo{author}{\bibfnamefont{A.~K.}
  \bibnamefont{{Tzioumis}}}, \bibinfo{journal}{MNRAS}
  \textbf{\bibinfo{volume}{346}}, \bibinfo{pages}{327} (\bibinfo{year}{2003}).

\end{thebibliography}
\end{document}